\documentclass[epj]{svjour}
\usepackage{graphicx}
\newcommand{\W}{8cm}

\begin{document}
\title{Numerical study of the temperature and porosity effects on the fracture propagation in a 2D network of elastic bonds.}
\author{Harold Auradou\inst{1,2}
\and Maria Zei \inst{1,3}
\and Elisabeth Bouchaud\inst{1}
}
\institute{Commissariat \`a l'Energie Atomique, DSM/DRECAM/Service de Physique et Chimie
des Surfaces et Interfaces, B\^at. 462, F-91191, Gif-sur-Yvette cedex, FRANCE. \and  Laboratoire Fluide, Automatique et Syst{\`e}mes Thermiques, UMR No. 7608, CNRS, Universit{\'e} Paris 6 and 11, B{\^a}timent 502, Universit{\'e} Paris Sud, 91405 Orsay Cedex, France. \and Laboratoire d'{\'e}tude des Milieux Nanom{\'e}triques, Universit{\'e} d'Evry, B{\^a}timent des Sciences, rue du p{\`e}re Jarlan, 91025 Evry Cedex.}
\date{\today}
\abstract{ This article reports results concerning the
fracture of a $2d$ triangular lattice of atoms linked by springs.
The lattice is submitted to controlled strain tests and the influence of both porosity and temperature on failure is investigated.
The porosity is found on one hand to decrease the stiffness of the material but on the other hand it increases the deformation sustained prior to failure.
Temperature is shown to control the ductility due to the presence of cavities that grow and merge.
The rough surfaces resulting from the propagation of the crack exhibit
self-affine properties with a roughness exponent $\zeta = 0.59 \pm
0.07$ over a range of length scales which increases with
temperature. Large cavities also have rough walls which are found
to be fractal with a dimension, $D$, which evolves with the
distance from the crack tip. For large distances, $D$ is found to be
close to $1.5$, and close to $1.0$ for cavities just before their
coalescence with the main crack. }
\maketitle
\section{Introduction}
Many materials such as cement or rocks have mechanical properties which are greatly influenced by the presence of pre-existing defects such as microcracks and micro-porosity due to their elaboration process \cite{Phys-Asp-Fract}.
In order to understand the mechanical behavior of such materials, different numerical modelling are developed.\\
The most classical approach consists in using discretization schemes for the continuum description.
The favorite scheme in fracture and damage mechanics is the finite element method.
Yet network models constitute an alternative scheme which has been developed in order to simulate the effect of heterogenities on the fracture process.
Network models can be classified in three categories. The
first group  is composed of {\it scalar models}, which exploit the
similarity between the failure of a heterogeneous material
submitted to an external load and the breakdown of an array of
randomly distributed fuses \cite{dbl86,hs03}. The results
can be used as a very interesting guideline, but in order
to compare theory with experiments, it is inevitable to consider
the vectorial nature of elasticity. In {\it the central force}
model \cite{pz01}, the bonds are springs which can freely rotate
around the site. The site, which will be called ``atom'' in the
following, undergoes displacements under the action of the local
forces acting on it. The third model, which will not be
considered here, is the {\it beam model} \cite{hhr89} which
contains full bond bending elasticity. In this case the elastic
energy of the beam is the sum of the elongation, shear and
flexural energies.
This contrasts with the spring model for which only the elongation of the bonds leads to their failure.\\
For all network approaches, bonds are supposed to model the
material at a "mesoscopic" level, and the aim is to investigate the
interrelation between disorder and properties of the network -
such as fracture stress or strain, and damage spreading. The
surprising result is that properties of the network are related to the system size by scaling laws involving non trivial exponents, independent of the precise distribution,
and of the microscopic aspects of the considered model \cite{ahhr89}.\\
Scaling is also observed on rough fracture surfaces for a large
variety of materials \cite{rev1,rev2} (from rocks \cite{psj92} to
wood \cite{mslv98} through metallic alloys \cite{blp90} and
glasses\cite{dag97,cpb03}), which can be described as self-affine
structures. Self-affinity \cite{book-feder88} means that a profile
extracted from such a surface, described by a heights
distribution $z(x)$, where $x$ is a Cartesian coordinate along the profile, remains statistically invariant under
the scale transformation $z(\lambda x)=\lambda^\zeta z(x)$, where
the Hurst or roughness exponent $\zeta$ characterizes the
roughness of the surface. The fluctuation of the surface heights
over a length $L$ is given by $\sigma_z(L)=\ell \left(L/\ell
\right)^\zeta$. Here $\ell$ is the topothesy, defined as the
horizontal distance over which fluctuations in height have a RMS
slope of one \cite{psj92}. For $3d$ fracture surfaces,
experimental values of $\zeta$ are found to be close to $0.8$, for
most materials \cite{psj92,mslv98,blp90,dag97,cpb03},
with the exception of some materials displaying intergranular
fractures, such as sandstone rocks, where $\zeta \sim 0.5$ \cite{bah98}. The exponent 0.5 was also
measured on glasses \cite{dag97} and on metallic alloys
\cite{ebn95,dag96} at length scales smaller
than the length scales at which the 0.8 exponent is observed \cite{rev1,rev2}.\\
Experiments conducted on two-dimensional samples reported somewhat smallest self-affine exponent; $0.6\pm0.1$ for paper \cite{san03} and $0.68\pm0.04$ for the fracture of wood, when the crack propagates along the fibers\cite{emhr94}.\\
In this paper, we present simulations of a mode I macro crack initiated by a notch
growing in a bidimensional porous material. The model is precisely
described in Section II. The initial pores are defined as regions
of the sample where bonds are missing.
In our model, the temperature of the network is controlled and its effect on the macroscopic mechanical behaviour of the system is studied in Section III.
In this section, the stress-strain curves corresponding to
a samples with no disorder and with  $30\%$ porosity are compared
for two values of the temperature.
It is shown that under an increasing strain, the pores will grow into cavities, and merge with each other and
with the main crack. Fracture hence proceeds by voids growth and
coalescence. The size and the density of the cavities is influence by the temperature.
At low temperature, the stress concentration due to the initial notch dominate the junction of cavities which are most likely collinear and located in its vicinity while at high temperature, the cavities spread over the whole materials.
This has strong consequence on the failure mechanisum: at low temperature, the material fractures in a  brittle way, while at
high temperature, it exhibits a ductile behaviour.
Once the porous samples are broken, we study the resulting
rough profiles, which are, like for real cracks, self-affine with
a roughness exponent $\zeta = 0.59 \pm 0.07$ that is independent of temperature. The
results of the analysis of the morphology of both the fracture
profiles and the cavities during their growth prior to failure are
presented in Section IV. Finally, Section V is devoted to
discussion.

\section{The model}

\begin{figure}
\includegraphics[width=\W]{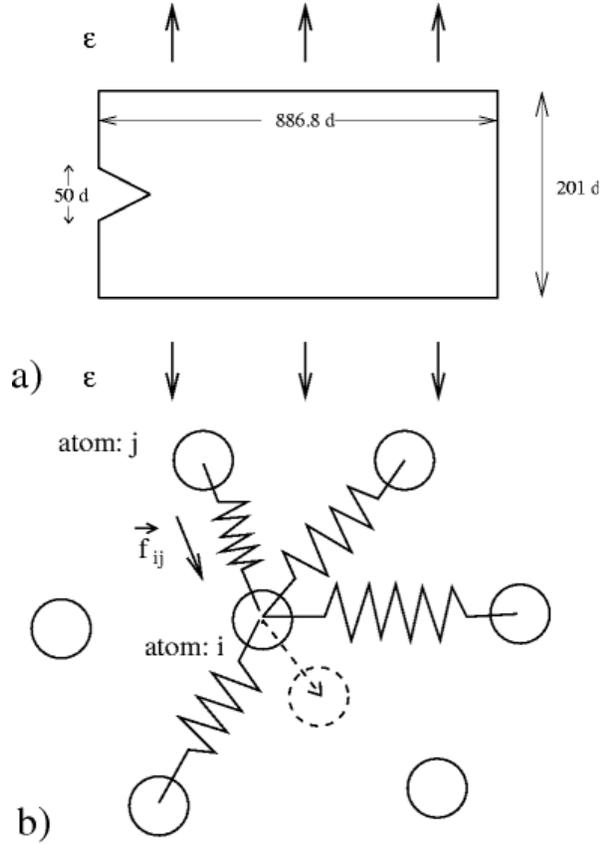}
\caption{Illustration of the elastic bond network used in the simulations. $a)$ shows the specimen submitted to an external uniaxial tension . $b)$ is a detailed view of the material consisting of atoms initialy placed on a triangular lattice and connected by identical elastic springs. Note that a fixed density of springs is removed before application of the load.}
\label{fig:figure1}
\end{figure}

The model consists in a $2d$ triangular lattice with nearest-neighbour
interactions (see Fig.\ref{fig:figure1}) that break as soon as the mutual distance becomes
larger than a prescribed threshold.
More precisely, by noting ${\vec r}_i$ the position of the $i^{th}$ ``atom'', the force ${\vec f}_{ij}$ due to the interaction with the
$j^{th}$ particle can be written as follows:
\begin{equation}
 {\vec f}_{ij}  =  F(|{\vec r}_i - {\vec r}_j|)
                   \frac{{\vec r}_i - {\vec r}_j}
                   { |{\vec r}_i - {\vec r}_j|}
\end{equation}
where $| \cdot |$ is the modulus and $F(u)$ is a scalar function
defining the force law. Here, we have chosen $F(u)$ to be a linear
function of the distance $u$ between atoms: $F(u) = -\alpha (u-d)$
(harmonic potential). The parameters are fixed in such a way that
$u=d$ is the equilibrium position and $\alpha$ is the spring
constant. In order to explicitly eliminate the irrelevant
parameters, we suitably rescale the spatial variables as well
time: in this way, both $\alpha$ and $d$ can be fixed to unity in
all that follows. Accordingly, all the quantities defined in this
paper are dimensionless.\\
In this work the size of the network is kept constant and is made of $68608$ triangular bonds. Due to the orientation of the lattice with
respect to the network (see Fig.\ref{fig:figure1}), its sizes in unit of atoms distance $d$ is $886.8$ for length, and $201$ for width.\\
As far as the sample is concerned, $c$ denotes the fraction of
initially missing bonds.  $c=0$ thus corresponds to a perfectly
homogeneous medium. Note that $c = c_p \equiv 0.653$ corresponds
to the ordinary percolation threshold: for $c> c_p$ so many bonds
are missing that the lattice is no longer macroscopically
connected \cite{stauffer}. Moreover, in the case of central
forces, there is a second threshold, the so-called
rigidity-percolation threshold ($c_r = 0.3398$ \cite{fs84,jt96})
above which the lattice although connected has zero Young modulus.
In what follows, the fraction of missing bonds is set to $c=0.3$.\\
A triangular notch of sides $50$ atoms is carved at the left side of the lattice to act as a stress concentrator and
force a main crack to propagate from the notch tip, along the $x$ direction (see Fig.\ \ref{fig:figure1}). The lattice is then
submitted to a controlled strain which acts vertically along the upper and lower sides of the sample,
 to which fixed boundary conditions are imposed, while free boundary conditions are chosen along the right and left borders.\\
The application of an external strain which gradually increases by
small steps of size $\delta \epsilon = 0.000725$ results in a
deformation of each spring, hence into atoms motion. Between two
successive increases of the strain, the new positions of the atoms
are computed. The first step of the calculation consists in
determining, for each atom, the force applied by its neighbours is
computed, and the various components are added to get the total
force acting on the considered atom. Newton's equation
\cite{fls63} is then solved for each atom $i$ (coordinates $\vec
r_i(t)$ ; velocity $\dot{\vec r}_i(t)$). For this purpose, we use
the leap-frog algorithm \cite{fls63,at87}, which is a modified
version of the Verlet algorithm.
This algorithm uses positions and accelerations at time $t$ and positions at time $t+\delta t$ to predict the positions at time $t+\delta t$,
where $\delta t$ is the integration step, set to the value $10^{-2}$. This step is repeated $N$ times before a new $\delta \epsilon$ increase
of the strain is imposed.\\
A bond breaks when it reaches a critical length $d^*$ which is set to the uniform value $1.1$. The fracture of a bond transforms its
potential energy into kinetic energy, which travels all over the lattice. A local dissipation, {\it{i.e.}} a force term
$-\gamma \dot{\vec r}_i$, is added along the left and right boundaries (Fig.\ \ref{fig:figure1}) where we expect the coupling with
the external world to be more efficient in removing kinetic energy from the medium. In the present work, complete damping, {\it i.e.}
$\gamma = 1$ is imposed.\\
A close look at the amount of kinetic energy present in the system
prior to any strain increase reveals fluctuation of the order of
$10\%$ with an average, $\langle E \rangle$ constant over the
whole range of strain including the loading and the failure parts
of the test. The parameter that controls the amount of kinetic
energy present in the network is the number $N$ of iterations used
to determine atomic positions. A decrease in $N$ results in an
increase of the amount of kinetic energy remaining in the network.
The latter is used to define a reduced equivalent temperature
\begin{equation}
T^*=\frac{\langle E \rangle}{\epsilon_c}
\end{equation}
where $\epsilon_c = 0.5 \alpha (d-d^*)^2$ is the energy needed to
break a single bond (under our conditions $\epsilon_c=0.005$). The
reduced equivalent temperature can be seen as the number of bonds
that the remaining kinetic energy might break
if it was not diluted in the network.\\
In the present work, two different values of $N$ are used:
$N=10^{5}$ and $N=10^6$, which lead respectively to reduced
equivalent temperatures $T^*=80$ and $T^*=8$. Before discussing
the quantitative results concerning the structure of damage and
the roughness of the fracture profiles, let us here briefly
illustrate the phenomenology that can be observed for the two
values of $T^*$.

\section{Macroscopic mechanical properties}
\label{sec:macro} 
\begin{figure}
\includegraphics[width=\W]{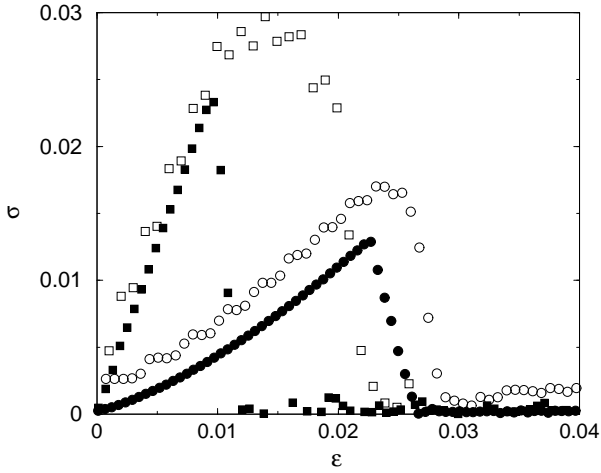}
\caption{Stress-strain evolution of the system. Circles and squares correspond
respectively to $c=0.3$ and $c=0$ (non-porous material).
Filled and empty symbols correspond respectively to $T^*=8$ and $80$.}
\label{fig:figure2}
\end{figure}

\begin{figure}
\includegraphics[width=\W]{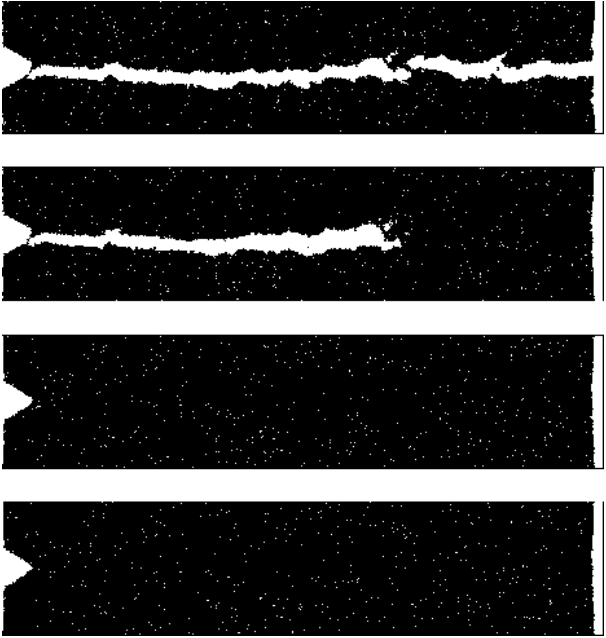}
\caption{Breakdown of the atoms network under strain test performed at the temperature
$T^*=8$. From bottom to top the strain is respectively $\epsilon=0.0223$, $0.0224$, $0.0225$
and $0.0282$.}
\label{fig:figure3}
\end{figure}

\begin{figure}
\includegraphics[width=\W]{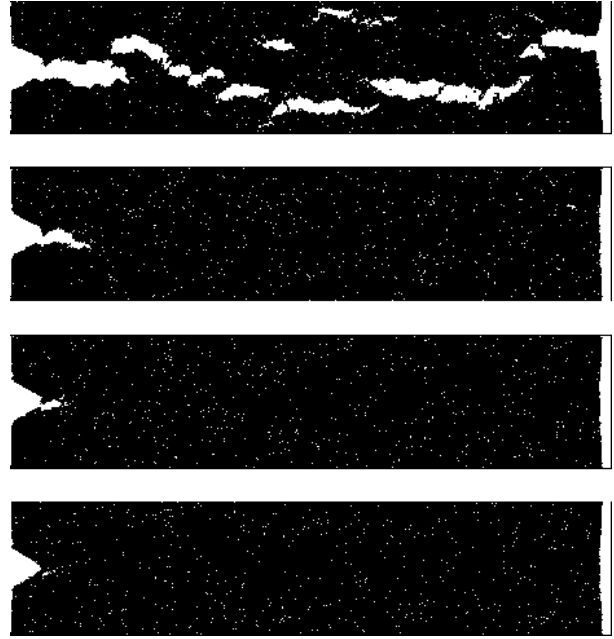}
\caption{Breakdown under tension of a network kept at the temperature $T^*=80$. From bottom to top the strain is respectively $\epsilon=0.0217$,
$0.0231$, $0.0225$ and $0.0289$.}
\label{fig:figure4}
\end{figure}

\begin{figure}
\includegraphics[width=\W]{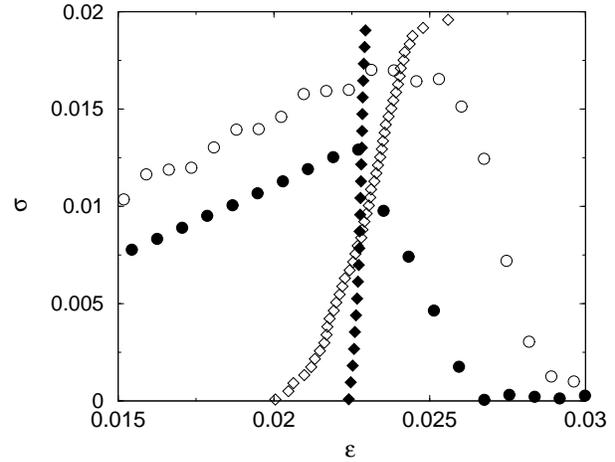}
\caption{Filled and unfilled symbols correspond respectively to
 $T^*=8$ and $80$.
Circles show the stress-strain variation for a porosity $c=0.3$.
Diamonds show the number of broken bonds verticaly normalised to fit the plot.}
\label{fig:figure5}
\end{figure}

Let us first examine the stress response at the
two different reduced equivalent temperatures. Figure
\ref{fig:figure2} shows the variation of the stress as a function
of strain for two sets of simulations performed at $T^*=80$ and
$T^*=8$, and for two different materials. The first material is
initially intact ($c=0$), meaning that no bonds were removed. From
the second one,
$30\%$ of the springs were removed at random ($c=0.3$).\\
After a first stage where the system gets easily deformed $-$ e.g.
for strains smaller than $0.005$ $-$, the stress-strain curves all
exhibit a linear behaviour. The stiffness decreases when bonds are
removed, from $2.5\pm0.2$ for
the intact material to $0.70\pm 0.04$ when $c=0.3$. This decrease does not seem to be temperature-dependent.\\
There is another major difference between the behaviours of the
two materials. For a given temperature, (see Tab.\
\ref{tab:table1}), the initially damaged material breaks at a
lower stress but sustains a higher
deformation. This is a typical ``quasi-brittle" behaviour, where toughening in an intrinsically brittle material is the result of damage
created ahead of the crack tip, which screens out the external field undergone by the main crack. \\
Let us now focus on the effect of the temperature. While the
stiffness is only a function of the density of remaining springs,
the maximum strain reached before the onset of crack propagation
increases with the temperature, as shown in Tab. \ref{tab:table1}.
At low temperature, a sharp decrease of the stress is observed
after the critical strain is reached. The strain-stress curve is
more rounded for a larger temperature. This effect reflects the
presence of damage ahead of the crack tip, as can been seen in
Figures \ref{fig:figure3} and \ref{fig:figure4}. It is clear from
these figures, that a temperature increase results in an increase
of the number of damage cavities. This can be seen on the dynamics
of bond failures: as shown in Fig.\ \ref{fig:figure5}, bonds start
breaking at a lower strain when the temperature is increased. For
$T^*=8$, the number of broken bonds as a function of strain almost
follows a step function, and increases abruptly when the crack
starts to propagate. This distribution broadens when the
temperature is increased up to $80$, showing that some of the
bonds are broken before the main crack propagation. Despite this
change in the shape of the distribution of the number of broken
bonds, which has a strong influence on the macroscopic mechanical
property of the network, the total number of broken bonds changes
only slightly, from $209$ for $T^*=8$ to $241$ for $T^*=80$, which
only represents approximately $0.15 \%$ of
the springs.\\
The other striking difference occurring when the temperature is
increased is an increase in the vertical shift of the
stress-strain curve. In fact, a linear fit of the data indicates
that the strain-stress curves do not pass through $0$. This
indicates that an excess of stress is present within the material.
This quantity is independent of the disorder and evolves from
$2.10^{-4}$ for $T^*=8$ to $2.10^{-3}$ for $T^*=80$. Note that stress
is applied via the forces acting on the surface atoms: a positive
stress excess thus indicates a force acting from the bulk toward
the outside and comes from the energy flux going from the network,
at temperature $T^*$, to its "cold" sides where complete damping
of the energy is imposed.
Moreover, as for a perfect gas, the stress acting on the sides is proportional to temperature.\\

This section points out that changes in the network porosity and temperature greatly influence its macroscopic properties.
The porosity eases the creation of damage cavities, the density of which is shown to be dependent on temperature.
At low temperature, the cavities are more likely ahead the crack tip; in the region where the stress is concentrated.
When the temperature rises, cavities spread over the network and the crack propagates in the damaged material by meandering from one cavity to another.
This phenomenon has a strong effect on the maximum strains that can be sustained by the structure.
The deviation from the main direction of propagation results, after failure, in rough fracture profiles. The next section is devoted to the analysis of their statistical properties.\\

\section{Self-affine properties of the fracture lines}
\label{sec:self-affine} 
\begin{figure}
\includegraphics[width=\W]{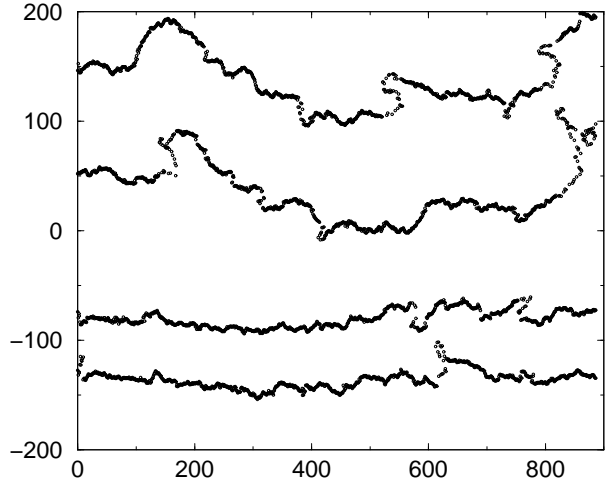}
\caption{Fractures profiles obtained after the breakdown of the lattice with $c=0.3$. The two top curves correspond to the top and bottom profiles obtained for $T^*=80$. The two lowest profiles are for $T^*=8$.}
\label{fig:figure6}
\end{figure}

\begin{figure}
\includegraphics[width=\W]{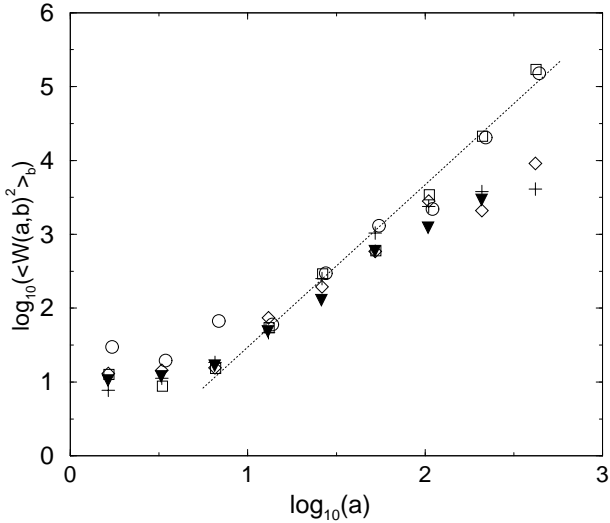}
\caption{$log_{10} \langle |W_{[y]}(a,b)|^2\rangle_b$ as function of $log_{10}(a)$ where $a$ is in unit of $d$. Circles and squares show the result of the analyse of the top and bottom profiles after failure at $T^*=80$ while diamonds and crosses are obtained when top and bottom failure profiles obtained at $T^*=8$ are considered. The filled triangles correspond to the average of the analyse of the two sides of the cavity at stage $(3)$ displays in Fig. \ref{fig:figure9}. The dotted lines has a slope of $2 \zeta + 1 = 2.2$ corresponding to a self-affine exponent $\zeta=0.6$. These results where shifted verticaly for convinience.}
\label{fig:figure7}
\end{figure}

\begin{figure}
\includegraphics[width=\W]{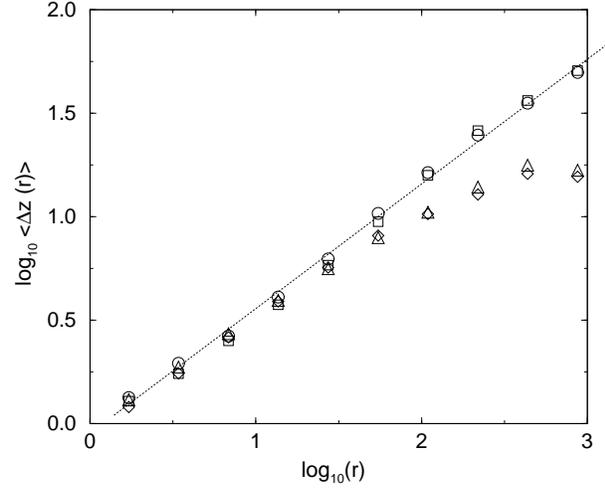}
\caption{$log_{10}(\langle \Delta z (r) \rangle)$ as function of $log_{10}(r)$. Circles and squares show the result of the analyse of the top and bottom profiles after failure at $T^*=80$ while diamonds and triangles are obtained when top and bottom failure profiles obtained at $T^*=8$ are considered. These results where shifted verticaly for convinience. The dotted line has a slope of $0.6$ .}
\label{fig:figure8}
\end{figure}

After each mechanical test, the positions
of atoms belonging to the two fracture lines are recorded. Figure
\ref{fig:figure6} shows the four profiles obtained from the two
tests performed at $T^*=8$ and $80$ on the porous material. In the
past years, various methods have been developed to measure the
roughness exponent of self-affine structures. In this paper, two
independent methods are considered namely the average wavelet
coefficient (AWC) analysis \cite{simonsen98} and
the min-max method \cite{book-feder88}.\\
In the case of the AWC analysis the one-dimensional line $z(x)$ is transformed into the wavelet domain as
\begin{equation}
W_{[y]}(a,b) = \frac{1}{\sqrt{a}} \int \psi^*_{a,b}(x) z(x) dx,
\end{equation}
where $\psi_{a,b}(x)$ is obtained from the {\it analyzing wavelet} $\psi$ (in our case a Daubechies
wavelet \cite{Daubechies}), via rescaling and translation, $\psi_{a,b} (x) = \psi((x-b)/a)$.
The AWC measures the average ``energy'' present in the profile at a given scale,
defined as the arithmetic average of $|W_{[y]}(a,b)|^2$ over all possible
locations $b$, and for a statistically self-affine profile with
exponent $\zeta$, it scales as: $\langle |W_{[y]}(a,b)|^2\rangle_b \sim
a^{2\zeta + 1}$.\\
For the second method, the profile of length $L$ is divided into
windows of width $r$. The linear trend of the line is then
subtracted from the profile for each window. The difference
$\Delta z(r)$ between the maximum and minimum height are computed
on each window and then averaged other all possible windows. For a
self-affine profile, a power law behavior is expected :
\begin{equation}
\langle  \Delta z(r) \rangle \propto r^\zeta
\end{equation}
For both methods, the self-affine scaling invariance will be revealed by data aligned along a straight line on a log-log
plot, with a slope which provides an estimate of $\zeta$.
Figures \ref{fig:figure7} and \ref{fig:figure8} shows log-log
plots of the results of the AWC and the min-max methods
respectively, for the four profiles considered. A self-affine
domain can be defined in each case and a self-affine exponent can be measured.
In the case of the wavelet analysis (Figure \ref{fig:figure7}),
$\zeta$ is found close to $0.60\pm0.07$ for $T^*=80$ and to
$0.55\pm0.1$ for $T^*=8$. For the min-max method (Figure
\ref{fig:figure8}), a linear fit indicates that $\zeta = 0.65\pm
0.1$ for $T^*=80$ and $ \zeta = 0.55\pm0.02$ for $T^*=8$. The
self-affine exponent characterizing the geometry of the profiles
may appear to depend slightly on the temperature, with a slight
increase when the temperature rises from $8$ to $80$. However, the
scaling domain is quite restricted (especially when the AWC method
is used), and the difference lies within error bars. When averaged
over the imposed temperature, the self-affine exponent is found to
be close to $0.59\pm0.07$. The difference in the lower cut-off
revealed by the two methods may be
attributed to the presence of overhangs on the profiles (see
Figure \ref{fig:figure6}), which are not included in the AWC
description, as discussed in \cite{dakh04}. Contrary to the value
of the exponent, the self-affine correlation length, defined as
the upper cutoff of the power-law domain,
 appears to be temperature-dependant and is found to be close to $100$ atoms spacing for $T^*=8$, while for $T^*=80$
 it overpasses the system size ($886$ interatomic spacing).\\

 In this section, we have pointed out that despite the ductility enhancement observed on the macroscopic mechanical properties when the temperature is raised, the resulting post mortem profiles have a self-affine roughness characterized by an exponent $\zeta =0.59\pm0.07$, independent of the temperature.
Yet, the temperature has a strong influence on the crossover length which separates the self-affine regime observed at small scales and the euclidean behavior displayed at large scales.
At low temperature, the growth of cavities is a consequence of the disordered structure of the sample, in a region close enough to the main crack tip for the stress to be high enough. Cavities nucleate from missing "atoms" in this region, and the process zone remains in the vicinity of the crack tip. The fracture profiles which result from the coalescence of the macro crack with the cavities have thus an amplitude which is limited by the lateral extension of the process zone. \\
On the other hand, when the temperature rises, the excess of stress due to the undamped kinetic energy (see Section \ref{sec:macro}) becomes non negligible compared to the stress created by the notch, and cavities are created everywhere in the lattice. In this case, the macro crack meanders through the whole network, and the amplitude of the post mortem profiles is larger.\\
Recently experimental and numerical observations of crack propagation in damaged materials suggested the existence of two self-affine domains \cite{rev2}.
At the scale of the cavity, the surface is characterized by an exponent $\zeta \sim 0.5$ while a larger exponent, $\zeta \sim 0.8$, is observed at the scale of the "superstructure" resulting from the coalescence of these cavities.
The next section is devoted to the quantitative analysis of the morphology of a single cavity.

\section{Structure of the damage zone}
\begin{figure}
\includegraphics[width=\W]{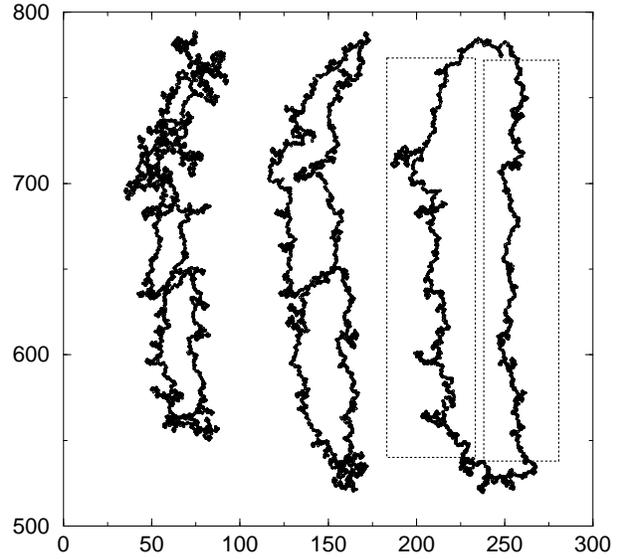}
\caption{From the left to the right, pictures of the same cavity at stages $(1)$, $(2)$ and $(3)$ of the test performed on the porous material at temperature $80$. The distance from the crack tip is respectively : $393$, $359$ and $355$ atoms spacing. The boxes indicate the left and right sides of the cavity at stage $(3)$.}
\label{fig:figure9}
\end{figure}

\begin{figure}
\includegraphics[width=\W]{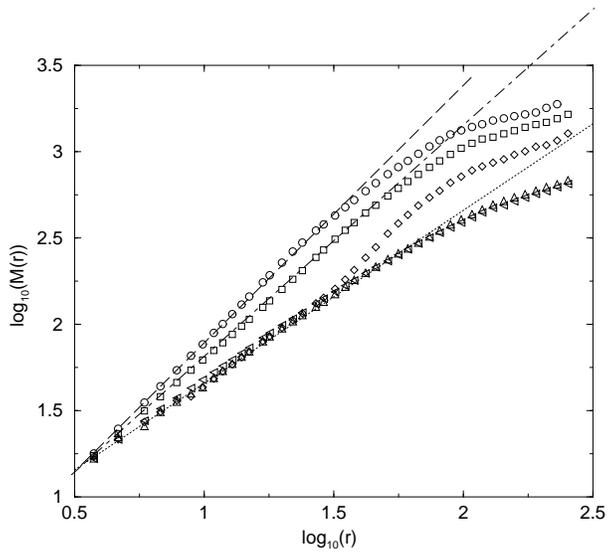}
\caption{Log log representation of the average mass $M(r)$ measured within circle of radius $r$ applied to the three cavities displayed in
figure \ref{fig:figure9}. Cirlcle, squares and diamonds are respectively for cavities $1$, $2$ and $3$. Triangles up and left are for the right and left sides of the cavity $(3)$ displayed in Fig. \ref{fig:figure9}. The long dashed, the dot dashed and the dahed lines have respectively a slope of $1.5$, $1.35$ and $1$.}
\label{fig:figure10}
\end{figure}

Figure \ref{fig:figure4} shows clearly that the morphology of the
crack profiles is influenced by the presence of cavities. In order
to describe quantitatively their evolution, we focus our attention
on one of the largest cavities. Figure \ref{fig:figure9} shows the
positions of atoms belonging to the external contour of the cavity
for three different values of the strain, during crack propagation.
Note that the total number of atoms, $1600$, belonging to this
contour remains unchanged through these three stages, and that the
first contour (stage $(1)$) is already the result of the
coalescence of smaller cavities. We clearly see on this figure
that as the crack tip gets closer to the cavity, the latter is
more open and elongated. Moreover, when the distance from the
crack tip is important, the contour shows meanders, the importance
of which decreases
as the crack tip gets closer. In order to describe the tortuosity of the contours and their possible scale invariance properties,
 the average mass method has been selected \cite{ahr01}.\\
This method is very similar to the box counting method and
consists in computing the number of atoms, $N(r)$ located within a
 circle of radius $r$ with its center located on one of the atoms of the contour.
 The average of $N(r)$ over all possible
 circle centers provides $\langle N(r) \rangle$. Figure \ref{fig:figure10} displays the evolution of
 $log_{10}(\langle N(r) \rangle)$ with respect to  $log_{10}(r)$, for the various contours.
 For a fractal contour, $\langle N(r) \rangle$ should increase with $r$ following a power law,
 $\langle N(r) \rangle \propto r^D$, where $D$ is the fractal dimension.
 For a smooth, Euclidean line, $D=1$, while for a line filling completely the plane, $D=2$.\\
Let us first focus on the behaviour of the contour of the cavity
at stages $(1)$ and $(2)$. For these two stages, the cavity
displays a fractal geometry over a domain of length scales
spanning from the atom spacing, $d$, up to approximately $50\ d$.
Its fractal dimension decreases from $D=1.5$ (stage $(1)$) to
$1.35$ (stage $(2)$). For stage $(3)$, the average mass displays a
more complex behaviour: for length scales smaller than $50\ d$,
the contour has a fractal dimension of $1$ but for larger length
scales, the fractal dimension seems to increase. In order to
understand this behaviour, we have analyzed separately the left
and ride sides of the cavity normal to the external load (Figure
\ref{fig:figure9}).
As shown in Figure \ref{fig:figure10}, the two sides are characterized by a fractal dimension $D=1$ and no abrupt change is detected.\\
The analysis of the contour of the cavity indicates that it is
indeed fractal, with a fractal dimension which decreases
continuously when the distance between crack tip and cavity
decreases. For large distances, the fractal dimension is found
close to $1.5$, but just before junction between the cavity and the main crack, the contour of the cavity has a fractal dimension of $1$.\\
It must be remembered however that the fractal dimension of a self
affine function is not uniquely defined:
it strongly depends on the range of length scales considered as well as on the method used.
As pointed out in the introduction, the height fluctuations of a self-affine profile is characterized by two parameters:
the self-affine exponent $\zeta$ and the topothesy $l$ which is the scale at which the slope of the profile is of the order of unity.
Above $l$, the fractal dimension is equal to $1$ for a $2d$ profiles. At smaller length scales, the dimension will depend on the method
used and is $D=2-\zeta$ for the average mass method. Because of the fact that in the present work the topothesy of the two sides of the cavity is less than the atom spacing the average mass method is not an appropriate method to analyse possible self-affine nature of the sides of the cavity.\\
As mentioned in Sec. \ref{sec:self-affine}, a more appropriate
tool to describe the self-affine nature of the profiles is the AWC
method. The latter has been applied to the two sides of the cavity
(See Fig. \ref{fig:figure7}), they display a self-affine
characteristic with an exponent, $\zeta \sim 0.6$, close to the value obtained for
the fracture profiles over length scales ranging from $6$ to $50$ atoms spacing.\\
 This section was devoted to the analyse of the morphology of a single cavity. Previous works suggested that cavities have rough walls with a self affine geometry characterized by a self affine exponent close to $0.5$ \cite{rev2}.
 The damage cavities obtained with our modelization is found to be self-affine with an exponent close to the one which
 characterizes the post mortem surface {\it i.e.} $0.6$. Yet, the self affine regime is observed over a narrow range of scale (less than one decade), making difficult any conclusion. Nevertheless, the structure of the walls of the cavity was found to be sensitive to the presence of the crack tip. This contrast with the scenario suggested in \cite{rev2} in which the surfaces created by the failure have a "quenched" geometry.

\section{Discussion}

Within the framework of a bidimensional numerical model, we have
examined crack propagation and damage spreading in a porous
material at two temperatures. We have shown that damage develops
more at high temperature, which results into a decrease of the
fracture strength and, correlatively, into an increase in
ductility. This increase in the elongation at failure results from
a screening of the external stress by damage. No plasticity is required in the model, which only involves bonds breaking
and atom rearrangements on a local scale. This behaviour is
similar to the one observed in quasi-brittle materials
 \cite{mslv98}-\cite{morelijf02}.

The crack morphology exhibits in both cases the same self-affine
roughness, with an exponent close to $0.6$ which agrees with measurement performed
on 2d materials \cite{san03,emhr94}. The structure of damage
at high temperature is also examined. Damage cavities are shown
to be fractal, with a fractal dimension which decreases from $1.5$
to $1$ prior to coalescence with the main crack. This change in the
fractal dimension is due to the increase of the local stress
generated by the closer vicinity of the crack tip during
propagation. This increase results in the coalescence of small
cavities and in atomic rearrangements of atoms on the cavity front
which can be interpreted as a partial "depinning". A similar
change in the morphology of cavities with their size was actually
observed recently in \cite{paun03}. The scaling properties of the
resulting crack is, in fine, due to the relative positions of
damage cavities with respect to each other rather than to their
structure, since they are no more fractal when they join the main
crack and become part of it. Further studies of the
inter-correlations of damage cavities for $2$- and $3$-dimensional
systems should lead to a better understanding of the still
mysterious morphology of fracture surfaces.

\begin{acknowledgement}
We are indebted to A. Politi, who is at the origin of the model used here. Many thanks also to R. Kalia, J-P Hulin, D. Bonamy and C. Guillot for their scientific support, and to Y. Meurdesoif and P. Kloos (CEA-Saclay Computer Science Division) for their technical support in the parallel simulations. HA is supported by the CNRS and ANDRA through the GdR
FORPRO (contribution No. $XXXX$) and the EHDRA (European Hot Dry Rock Association) and PNRH programs.
\end{acknowledgement}


\newpage
\begin{table}
\begin{tabular}{l|c|c|c|l|c|c|c}
      & $T^*$ & $80$ & $8$ &  & $T^*$ & $80$ & $8$\\
  \cline{2-4} \cline{6-8}
$c=0$ & $\epsilon_{\sigma_{max}}$ & $0.014$   & $0.0096$ & $c=0.3$  & $\epsilon_{\sigma_{max}}$ & $0.023$   & $0.022$ \\
      & $\sigma_{max}$   & $0.029$  & $0.024$  & & $\sigma_{max}$   & $ 0.017$  & $0.013$  \\
      & $\epsilon_{max}$ & $0.024$  & $0.012$         & & $\epsilon_{max}$ & $0.029$  & $0.026$\\
\end{tabular}
\caption{Mechanical properties of the system as a function of the temperature $T^*$
for two different fractions of missing springs $c=0.0$ and $c=0.3$. $\epsilon_{max}$, $\sigma_{max}$ and  $\epsilon_{\sigma_{max}}$
stand for the maximum strain (reached at the onset of complete failure), the maximum stress reached during the test, and
the corresponding strain.}
\label{tab:table1}
\end{table}

\end{document}